\documentclass[twocolumn]{bmcart}
\usepackage{amsthm,amsmath}
\usepackage{graphicx}
\RequirePackage{natbib}
\usepackage{color}

\definecolor{darkblue}{rgb}{0,0,0.5}
\definecolor{lila}{rgb}{0.3,0,0.3}
\definecolor{turq}{rgb}{0,0.1,0.4}
\definecolor{lightblue}{rgb}{0.7,0.7,0.9}
\usepackage{url} 

\usepackage[pdftex,
 colorlinks=true,
 backref=page,
 linkcolor=darkblue, 
 filecolor=red,
 citecolor=turq, 
 urlcolor=lila, 
 pdftitle={Single Molecule DNA Detection with an Atomic Vapor Notch Filter},
 pdfauthor={Denis Uhland and Torsten Rendler and Matthias Widmann and Sang-Yun Lee and Jorg Wrachtrup and Ilja Gerhardt},
 pdfsubject={},
 pdfkeywords={DNA Detection; Fluorescence Microscopy; Single Molecules; Atomic Filtering; Sodium Spectroscopy},
 pdfpagelabels=true, 
 breaklinks=false,
 plainpages=false,
 backref=false,
 bookmarks,
 bookmarksnumbered=true]{hyperref}
\usepackage[utf8]{inputenc} 
\startlocaldefs
\newcommand{\degree}{\ensuremath{^\circ}}

\endlocaldefs
\begin{document}
\begin{frontmatter}

\begin{fmbox}
\dochead{Research}
\title{Single Molecule DNA Detection\\ with an Atomic Vapor Notch Filter}

\author[
   addressref={aff1},                   
]{\inits{DU}\fnm{Denis} \snm{Uhland}}

\author[
   addressref={aff1},
]{\inits{TR}\fnm{Torsten} \snm{Rendler}}

\author[
   addressref={aff1},
]{\inits{MW}\fnm{Matthias} \snm{Widmann}}

\author[
   addressref={aff1},
]{\inits{SL}\fnm{Sang-Yun} \snm{Lee}}

\author[
   addressref={aff1,aff2},
]{\inits{JW}\fnm{J\"org} \snm{Wrachtrup}}

\author[
   addressref={aff1,aff2},
   email={i.gerhardt@fkf.mpg.de}
]{\inits{IG}\fnm{Ilja} \snm{Gerhardt}}

\address[id=aff1]{
  \orgname{3rd Physics Institute, University of Stuttgart and Stuttgart Research Center of Photonic Engineering (SCoPE) and IQST}, 
  \street{Pfaffenwaldring 57},                     %
  \postcode{70569}                                
  \city{Stuttgart},                              
  \cny{Germany}                                    
}
\address[id=aff2]{%
  \orgname{Max Planck Institute for Solid State Research},
  \street{Heisenbergstra\ss e 1},
  \postcode{70569},
  \city{Stuttgart},
  \cny{Germany}
}

\end{fmbox}

\begin{abstractbox}

\begin{abstract} 
The detection of single molecules has facilitated many advances in life- and material-sciences. Commonly, it founds on the fluorescence detection of single molecules, which are for example attached to the structures under study. For fluorescence microscopy and sensing the crucial parameters are the collection and detection efficiency, such that photons can be discriminated with low background from a labeled sample. Here we show a scheme for filtering the excitation light in the optical detection of single stranded labeled DNA molecules. We use the narrow-band filtering properties of a hot atomic vapor to filter the excitation light from the emitted fluorescence of a single emitter. The choice of atomic sodium allows for the use of fluorescent dyes, which are common in life-science. This scheme enables efficient photon detection, and a statistical analysis proves an enhancement of the optical signal of more than 15\% in a confocal and in a wide-field configuration.
\end{abstract}

\begin{keyword}
\kwd{DNA Detection}
\kwd{Fluorescence Microscopy}
\kwd{Single Molecules}
\kwd{Atomic Filtering}
\kwd{Sodium Spectroscopy}
\end{keyword}

\end{abstractbox}
%

\end{frontmatter}

\section*{Introduction}

The optical detection of single molecules~\cite{hirschfeld_ao_1976,moerner_prl_1989,orrit_prl_1990} has facilitated important progress in various fields of research. Especially in microbiology the detection, localization and tracking of tagged biomolecules~\cite{hell_oib_1994,betzig_s_2006,rust_nm_2006} or other relevant structures like DNA molecules~\cite{schafer_nature_1991,nie_science_1994,nie_arobabs_1997,weiss_science_1999} support the deep insights into the underlying composition and functionality of living cells. Commonly, the red-shifted fluorescence of single molecule labels is detected. The number and the emission rate of photons is commonly limited by various factors: The radiative decay rate of the emitter, its non-radiative decay channels, the extraction efficiency of light out of the sample under study, and the collection and detection efficiency of the utilized microscope. Under ambient conditions the overall number of extractable photons of organic fluorophors is limited, since a probabilistic photo-bleaching~\cite{gordon_pnas_2004} event stops any further data-acquisition. All interesting parameters, such as the localization accuracy, are limited due to the finite number of detected photons. This occurs usually in a shot-noise fashion, which scales as $1/\sqrt{N}$, where $N$ is the number of detected photons. From a technical standpoint mainly two key parameters can be optimized: The brightness of the probe used as a tag can be higher, or the collection~\cite{lee_np_2011} and detection efficiency of the microscope is enhanced. 

In the last decade there have been many efforts to optimize both experimental parameters. Various experiments with defect centers in diamond~\cite{jelezko_sm_2001,vlasov_nnano_2014} and semiconductor nano-crystals~\cite{biju_aabc_2008} with suppressed photo-bleaching were performed~\cite{evanko_nmeth_2008}. Other attempts have been established to increase the extraction efficiency out of the structure under study~\cite{lee_np_2011,jamali_rosi_2014}. Additionally, the detectors have been continuously improved, such that nowadays avalanche photon diodes (APDs) exhibit more than 70\%, and sensitive CCD-cameras more than 95\% quantum efficiency. Another way to enhance the overall detection efficiency for single emitters is an optimized filtering scheme, since the red-shifted fluorescence is detected. An ideal filter solely blocks the excitation light which is elastically scattered by the sample, and transmits all Stokes-shifted photons. The edge-steepness should be ideally a step-function, but is often limited by technical imperfections in usual dichroic mirrors and filters.

Also hot atomic vapors can allow for optical filtering. These are generally easy to handle in evacuated glass reference cells. This is an evacuated glass cylinder with optical windows, in which a small amount of alkali metals (hundreds of milligram) is present. Such cells can exhibit a large optical depth, and simultaneously ensure a few GHz spectral width. The optical rejection founds on Beer-Lamberts law and is further a function of the vapor density, which rises approximately exponentially with temperature. The spectral width is to a first approximation given by the Doppler broadening of an atomic vapor, in the range of a few GHz at ambient conditions up to a few hundred degree centigrade~\cite{weis_mathematica}. In atom optics experiments such filtering schemes are common and have been quantified for different alkali metals like rubidium~\cite{heifetz_oc_2004,weller_jpb_2012}. One common use is to filter the emission of one isotope of rubidium with another, an application which is commonly implemented in atomic clocks~\cite{camparo_pt_2007}. These filters are also suitable for other applications like Raman spectroscopy~\cite{pelletier_as_1992,chen_ol_1996,junxiong_ao_1995,lin_ol_2014}, and are intrinsically matched to atomic transitions.

Here we present the detection of single fluorescing molecules in a confocal~\cite{minsky_patent_1961} and a wide-field microscope~\cite{singer_science_1932}. We compare the filtering scheme between a high-end commercial filter and hot atomic sodium vapor. Unlike other atomic filters, this atom selection matches the visible range of many common dye systems which are used for biological labeling. The molecules under study are single stranded DNA molecules, labeled with a commonly used fluorescence dye (ATTO 590). Both microscopic schemes, confocal and wide-field, have their specific advantages and disadvantages. We are able to show that filtering with atomic vapor is able to facilitate an enhanced detection of the fluorescence which originates from a single molecule. Initial steps of such a study were presented for single molecules under cryogenic conditions earlier~\cite{siyushev_nature_2014}.

\section*{Experimental Configuration}

\begin{figure}[ht]
  \includegraphics[width=0.9\columnwidth]{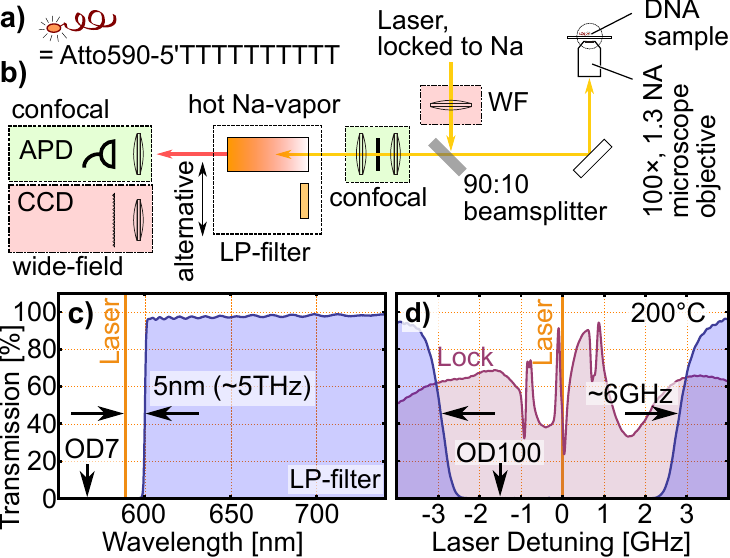}
  \caption{a) single stranded DNA under study, labeled at the 5'-end with the organic dye ``Atto590''. b) Confocal and wide-field microscope. The filter configuration can be changed in approx. 1~sec. WF=wide-field configuration; APD=avalanche photo diode; LP-filter=long-pass filter; CCD=charged coupled device, camera; c) laser emission and filter function of the commercial long pass filter. OD=optical density; d) laser and filter function of the atomic vapor cell at 200\degree C. The block band is approx. 6~GHz broad and shows an optical rejection of more than 6 orders of magnitude (measured). The calculated optical density is much higher (OD 100).}
  \label{fig:setup}
\end{figure}

The experimental configuration consists of a combined confocal and wide-field microscope (Fig.~\ref{fig:setup}a, b). Both experimental configurations are described below. Filtering is performed either with a commercial filter or with an atomic vapor cell.

\subsection*{Excitation Laser}

The excitation laser is a dye ring laser (699-21, Coherent), which can be locked to a sodium transition. The lock-signal is provided by Doppler-free dichroic atomic vapor laser lock (DAVLL, see e.g.~\cite{petelski_epjd_2003}) with a 140\degree C hot sodium vapor cell. For optimal optical rejection, the laser is locked to the D$_2$-line cross-over resonance, midways between the two $F=1$ and $F=2$ ground states of the D$_2$ line (see Fig.~\ref{fig:setup}d). This represents for further filtering purposes a preferred point, since both optical transitions add up due to their Doppler broadening. This zero-point does not represent the ``center of gravity'' of the unshifted sodium transition. The lock-signal is robust against external influences such as mechanical noise on the laser table. The dispersive lock signal is depicted in Fig.~\ref{fig:setup}d. For the rare case the laser jumps out of lock, the signal is monitored with the help of the pass-fail output of an oscilloscope (LeCroy). This controlled an optical shutter and blocks the laser if required to prevent damaging of the used single photon detector. From the dye laser a 20~m long optical single mode glass fiber guides the laser beam to the setup. The light is then filtered with a narrow-band band-pass filter (589$\pm$0.2~nm, Omega Optical), to suppress fiber fluorescence or background due to the broad-band fluorescence of the dye laser itself.

The base of the microscope is built around a commercial inverted microscope (Olympus IX71). It is used in both, a confocal, or a wide-field configuration.

\subsection*{Confocal Microscope}

For the confocal configuration, the collimated beam (diameter: 4.8~mm) is reflected into the microscope via a quartz wedge (10\% reflectivity). This is not realized by a commonly used dichroic beam-splitter to avoid additional spectral cut-off of the detected fluorescence signal. The light is focused onto and collected from the sample with a 100$\times$, 1.3~NA microscope objective (UPLANFL, Olympus). Detection is performed by focusing the resulting light onto a 50~$\mu$m pinhole. From there, the light is again 1:1 collimated and passes the filter configuration under study. The light is then focused onto a single photon counting module (SPCM-AQR-14, Excelitas). Addressing different locations on the sample and focusing is realized by sample scanning with a 3D-piezo actuator (P527.3CL, Physik Instrumente). For sample scanning, a pixel size of 100~nm was used. Typical integration times per pixel were 10~ms. The entire detection scheme is carefully optically shielded from the environment and in more detail discussed below. 

To acquire single molecule spectra, a flip-mirror is introduced into the confocal configuration to guide the light from the single photon detector to a cooled CCD spectrometer (Princeton Instruments, Acton, 300~mm, camera: ``Pixis''). An acquisition time of 30~sec. is used.

\subsection*{Wide-field Microscope}

In further experiments the configuration is changed to a wide-field microscope. The incident light is then focused into the back-focal plane of the microscope objective. This is checked by observing a collimated beam above the microscope. Due to the extension of the microscope enclosure, this focusing is not performed by a single lens, but with a pair of a 100~mm and a 50~mm achromatic lens. The illuminated area on the sample is about 25~$\mu$m in diameter. For imaging, a cooled CCD camera is used (Photometrics Cascade 512B) with a pixel size of 16$\times$16~$\mu$m$^2$ and 512$\times$512 pixels. Focusing on this camera is performed with a 250~mm achromatic lens, such that a single pixel had an extension of 115$\times$115~nm$^2$ on the sample. A common acquisition was performed with an exposure time of 10~sec and with no further internal gain.

\subsection*{The Sample}

The single stranded deoxyribonucleic acid (DNA) under study consists of 10 bases (5'-T\-T\-T\-T\-T\-T\-T\-T\-T\-T, see also Fig.~\ref{fig:setup}a) and is labeled on the 5'-end with Atto-590 (Thermo Scientific). This dye has been chosen to be a common dye label in micro-biology and the spectral match to atomic sodium vapor. The sample is produced by dissolving and diluting the dried DNA in sterile water. An aqueous polyvinyl alcohol (PVA) solution (3~mg/ml) is mixed to the DNA solution until a relative DNA concentration of 1:10$^{14}$-1:10$^{15}$ is reached. The solution is then spin-coated onto clean cover slides. For the presented experiments a concentration of 1:10$^{15}$ has been used, since single molecules are separated by several $\mu$m, whereas the 1:10$^{14}$ diluted sample shows a too dense concentration for automatic peak-find routines. The thickness of the sample has been calculated based on the assumed concentrations and is below one $\mu$m.

To ensure that the performed experiments are on the single molecule level, the typical blinking behavior of single molecules is confirmed in the wide-field configuration with millisecond integration times. Complementary, a recording of a confocally detected single molecule is shown in Fig.~\ref{fig:spectra}b. The single step bleaching around t=500~sec indicates that a single molecule was observed. This furthermore gives a reference for the possible acquisition times. Note, that no telegraph function or triplet blinking can be observed due to the long integration time.

\subsection*{Filtering the Excitation Light}

The optical filtering from the microscope is performed with two possible filter configurations: a commercial filter and an atomic sodium vapor cell.

\begin{figure}[ht]
  \includegraphics[width=0.9\columnwidth]{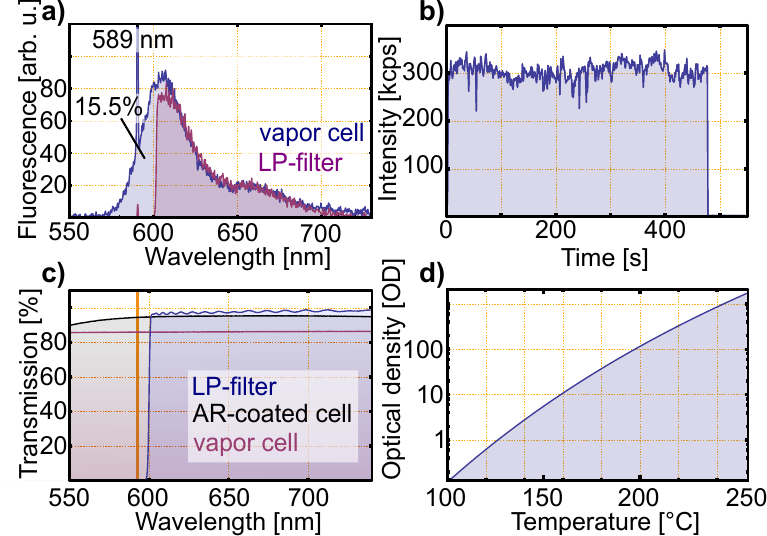}
  \caption{a) Normalized emission spectra of the dye Atto590 under illumination with laser light at 589~nm, filtered solely with a hot (200\degree C) atomic vapor cell. Purple: Same spectrum, filtered with the commercial long-pass filter. The integral enhancement between atomic and the commercial filter is 15.5\%. LP-filter=long-pass filter. b) Single step bleaching of individual molecules proves the single emitter nature. c) Transmission curves of the filter and different atomic vapor cells. d) Calculated optical density of the filter (100~mm optical length) against temperature.}
  \label{fig:spectra}
\end{figure}

The commercial filter (Semrock, FF01-593/LP-25) represents a very good choice for the excitation light and exhibits around 97\% transmission in its pass-band. A spectrum of the filter was recorded in a commercial absorption spectrometer (Perkin-Elmer, Lambda 16) and is depicted in Fig.~\ref{fig:setup}c and \ref{fig:spectra}c. Please note, that this is a linear representation, and the 50\% point is observed around 600~nm. This implies that the optical rejection of six to seven orders of magnitude will be only efficient around 594~nm or shorter. Since the filter is usable for 593~nm, it is likely possible to slightly tilt the filter for a blue shifted spectrum, closer to the excitation laser. By this, also the transmission of the filter might be altered, and therefore this option was not pursued.

Alternatively, the excitation light, scattered and reflected from the sample, is blocked solely by a hot atomic vapor cell. A simple calculation~\cite{weis_mathematica} of the six D$_2$ transitions ($3^2S_{1/2}\rightarrow 3^2P_{3/2}$, three from each ground state) allows to estimate the optical density for the 100~mm long cell (30~mm diameter) as a function of the temperature (Fig.~\ref{fig:spectra}d). This is an idealized picture, since any forward scattering, non-linear or saturation effects are not accounted for. Furthermore, the laser has to represent a delta-peak, locked to the cross-over resonance of the sodium D$_2$-line. The D$_2$-line is generally preferred for filtering, due to its by factor of two higher oscillator strength. Nevertheless, we measure a suppression of better than 6 orders of magnitude with the filter at a temperature of about 200\degree C. This is the operation temperature of the filter for all further experiments. The measured filter function of the sodium vapor is depicted in Fig.~\ref{fig:setup}d. The optical density of about 100 represents the calculated value.

Atomic sodium tends to diffuse into usual boro-silicate glass and darken the windows. Therefore, the cells were produced out of quartz glass. Other methods, such as special glasses, allow for a suppression of this darkening effect~\cite{laux_jopesi_1980,sakurai__1990}. For filtering purposes, a cell with anti-reflection coated windows was obtained (Triad technologies, Longmont, Colorado). Although unspecified, the supplied coating works well up to 200\degree C, and has superior transmission than the quartz glass alone. A different batch of cells was produced in house, but without the anti-reflection coating. The transmission spectra of the used cells, and the commercial filter is shown in Fig.~\ref{fig:spectra}c. With the resolution of the utilized absorption spectrometer, the GHz-wide suppression of the sodium light is not observed. 

To directly compare the two filter configurations in the microscope, both filters are mounted onto a small optical bread board, which slides on rails on the optical table. The filter configuration can therefore be switched within less than a second back and forth. The area between the filters was filled with an opaque material, such that the filter configuration can be changed during a running experiment when all detectors are on. Furthermore, the entire setup was optically shielded inside a black cardboard box. To ensure a well defined convection of air inside the box, it was designed such that a laminar flow was realized from bottom to top by small holes at the bottom.

\section*{Results}

\subsection*{Acquired spectra}
We first compare the different spectra, acquired from a single molecule in both filter configurations. When a single molecule was identified in the confocal microscope, the spectrometer was introduced and a spectrum was acquired. Fig.~\ref{fig:spectra}a shows the single molecule spectra in both filter configurations. The filtering with the atomic vapor shows the entire spectrum of the single molecule as it would be excited with a more blue wavelength. On the other hand, we find the spectrum acquired with the commercial 593~nm long-pass filter rising at around 600~nm. Interestingly, the spectrum acquired with the atomic filter shows contributions which are lower in wavelength than the excitation light. On a first sight, this violates energy conservation, but likely the more blue components are introduced by anti-Stokes processes. Furthermore, compared to the original spectrum, provided by the dye producer, we find the single molecule spectrum to be spectrally shifted to the blue part of the spectrum by about 5~nm. This is not untypical for organic dyes that the fluorescent properties critically depend on their chemical environment~\cite{kubin_jol_1983,arbeloa_cp_1989}.

When the quantum efficiency of the single photon detector is assumed as a fixed value in the range of 570-700~nm, we can directly compare the integral contribution of the dye spectra as proportional to the detected signal on a photo detector. By a simple integration, we find a higher signal with atomic filtering of about 15.5\%. This should be the effective enhancement of the single molecule signal on a detector when the atomic filtering is introduced.

The sodium vapor filter not only increases the overall signal from the molecule under study, but also introduced a higher laser background. Although the optical rejection is calculated to be much higher than with the commercial filter, we find an about 20 times increased laser background. There are two explanations for this finding: a) the filter does not have such a high rejection due to non-linearities or saturation effects. Or, b), the contribution of scattered and spectrally shifted light around the laser wavelength is larger than which is observed with a commercial filter. To address this problem, we performed simple measurements of the atomic filter with directly from the microscope reflected laser light. This resulted in a measured optical rejection of more than 6 orders of magnitude. If these measurements were performed far below the saturation intensity (9.4~mW/cm$^2$), we would be unable to determine the optical rejection of the filter due to the weak signal. Generally, the weak laser background contribution is not relevant for the acquisition of single molecule signals, since this simply adds a linear background. If laser power fluctuations can be circumvented, the signal to background (SBR) ratio is affected, but not necessarily the strength of the molecular signal on the background. In such a case, the background contribution can be simply subtracted. 

\subsection*{Data Analysis}

For the confocal and the wide-field imaging, data processing was performed as follows: Since the main goal was to compare the two filter configurations, an image with each filter was acquired with the exact same settings (excitation intensity, acquisition time, etc.). Both images were acquired directly after each other. It was furthermore ensured, that not only one filter configuration was acquired first, but usually a sequence of the commercial filter, the atomic filter, and the commercial filter again. This procedure ensures that photobleaching, which can only occur in the second of two acquired images, does not lead to a systematic error. Then, the two corresponding images were processed by an automated peak-find routine to identify the molecules. If one peak was laterally found within the same position ($\pm$ 3~pixel) in both images, they were considered as the same molecule. A lateral shift of the images between the both filter configurations was not observed. Each peak was then least-square fitted with a 2D-Gaussian (symmetric in $x$ and $y$) and the integral area below the curve, background level, and the lateral width were determined. This procedure results in two histograms (not shown), where the intensity values for \emph{all} acquired molecules are listed. Further, the derived values are compared. For the integral area, i.e.\ the emitted intensity from the molecule, we find a rise at a certain minimal power, which corresponds to the threshold-parameter of the peak-find routine.

This histogram might principally be used to characterize the enhancement (or suppression) of the detected light intensity by using the atomic filter cell. Unfortunately, the direct comparison would be misleading, if any parameters changed between the acquisition of different pairs of images. Therefore, each pair of molecules was directly analyzed in both filter configurations. The fitted integral amplitude was compared only between one molecule in both filter configurations at a time. Since the molecules tend to bleach and blink at ambient conditions, this required a statistical analysis. It is evident that each molecule can be found brighter or dimmer within a certain acquisition time. 

To estimate the background and its fluctuation, all images were analysed for their background level and fluctuations in an area where no molecules were present. Due to the fact that we can analyze many more molecules than images, the statistical fluctuation of this analysis is higher. With this data, a histogram with signal to background (SBR) and signal to noise ratios (SNRs) can be determined. This calculation was simplified and the ratio is simply calculated as the ratio of the average signal height to the standard deviation of the background without any molecules. The fluctuation on the molecules emission signal was not accounted for. All non-fluctuating background contribution can be simply subtracted from the original data. The analysis presented below is always a statistical comparison between both filter configurations and many co-localized single molecules.

\subsection*{Confocal Imaging}
\begin{figure}[ht]
  \includegraphics[width=0.9\columnwidth]{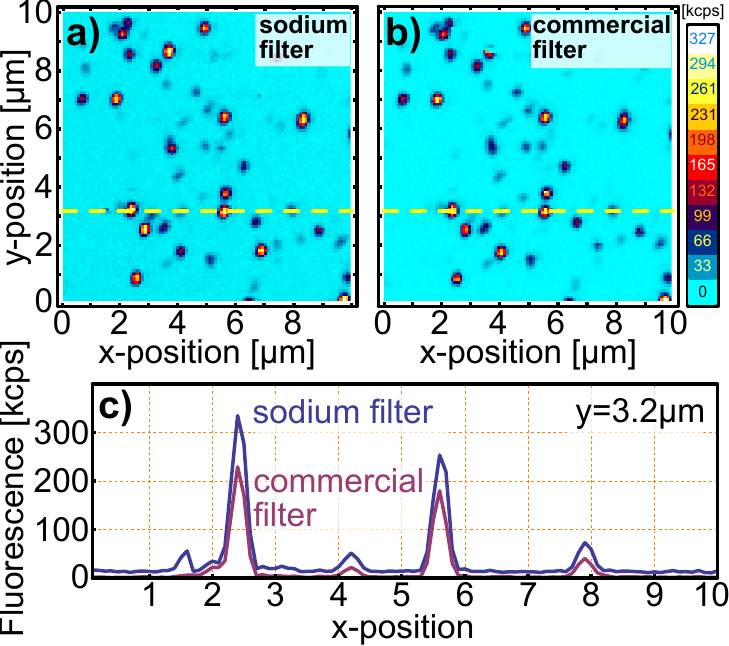}
  \caption{a) Confocal image of the sample, filtering with the atomic vapor cell (T=200\degree C). Pixel size is 100~nm. Integration time per pixel 10~ms. b) same with filtering by the commercial filter c) Line-cut of single molecules emission extracted from the above image (yellow dashed line).}
  \label{fig:confocal}
\end{figure}

Now we turn to the confocal configuration of the microscope. Initially, the experimental configuration was differently set up than shown in Fig.~\ref{fig:setup}b, and the filter configuration was placed between the microscope and the pinhole. Due to thermal scintillation of hot air, which originated from the hot vapor cell, this led to severe power fluctuations in acquiring single molecule signals with the atomic filter. Any scintillation affects the collimated beam and the image on the pinhole wanders, respectively blurs out. The observed power fluctuations are on the order of one magnitude under these conditions. Therefore, the confocal configuration was changed and the atomic filter was placed between the pinhole and the single photon detector. This configuration is also described in~\cite{lin_ol_2014}. To estimate the fluctuation of the wandering beam, we placed a camera at the location of the avalanche photo diode. At high speed (10~ms), a 71$\mu$m spot size ($1/e^2$) is observed, as if no thermal fluctuations are present. With an integration over 60~sec., a spot size of 82$\mu$m is observed. The particular configuration of the vapor cell behind the pinhole does not reduce the spatial resolution compared to the commercial filter. For the vapor cell we see a comparable spatial resolution of 291~nm compared to 294~nm with the commercial filter. 

A raw confocal image of the single molecule sample in both filter configurations is shown in Fig.~\ref{fig:confocal}a and b. Single step bleaching of a molecule is observed in the commercial filter configuration ($x,y$=4,8.5~$\mu$m). Visually, both images are comparable, but a higher background contribution is observed in atomic filtering around 15~kcps vs.\ very stable 2~kcps with the commercial filter. Furthermore, the noise level is affected, and the background fluctuation is increased by a factor of 15 with the atomic filter. Subsequently, we estimate the signal to noise ratio to be by an order of magnitute larger with the commercial filter (SNR=800!). This is differently than what has been observed in the experiments under cryogenic conditions~\cite{siyushev_nature_2014}. The increased power levels at ambient conditions ($\mu$W instead of nW) seem to have an influence on the signal to noise ratio. It underlines the hypothesis, that saturation effects in the vapor are diminishing the filter performance.

The line cuts shown in Fig.~\ref{fig:confocal}c illustrate the background level. The image background acquired with atomic filtering is higher by a factor of about 7-8 based on a laser-power of 1~$\mu$W into the microscope. This is fully consistent with the measurements of the optical density. The commercial filter shows an optical density of about seven, whereas the filter cell was determined to show only six orders of magnitude optical suppression.

\begin{figure}[ht]
  \includegraphics[width=0.9\columnwidth]{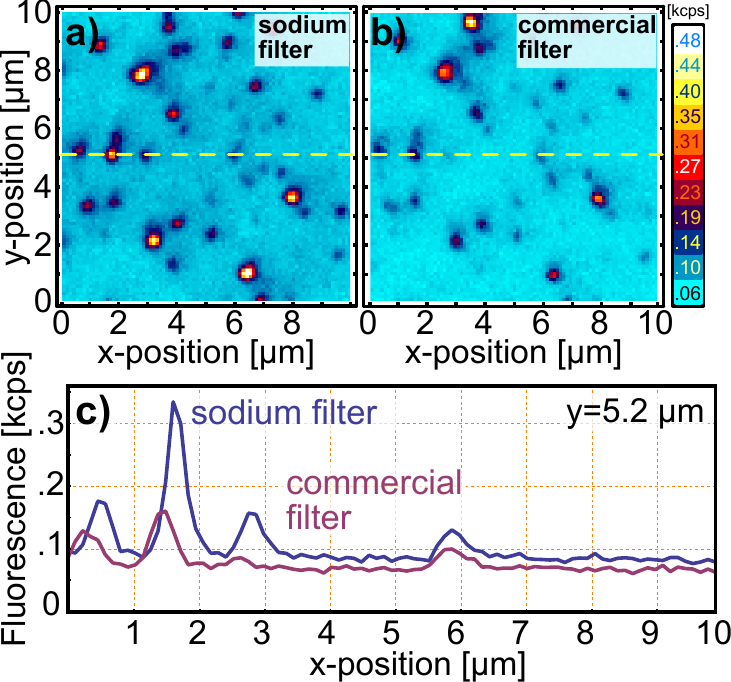}
  \caption{a) Wide-field image of the DNA sample, atomic filtering (T=200\degree C). Acquisition time per frame 10~s. b) same with filtering by the commercial filter. c) line-cut of a single molecule (yellow dashed line). Due to the wide-field configuration and the scintillation, the lateral extension of the molecules tend to be larger than in the confocal case.}
  \label{fig:widefield}
\end{figure}

\begin{figure}[h]
  \includegraphics[width=0.9\columnwidth]{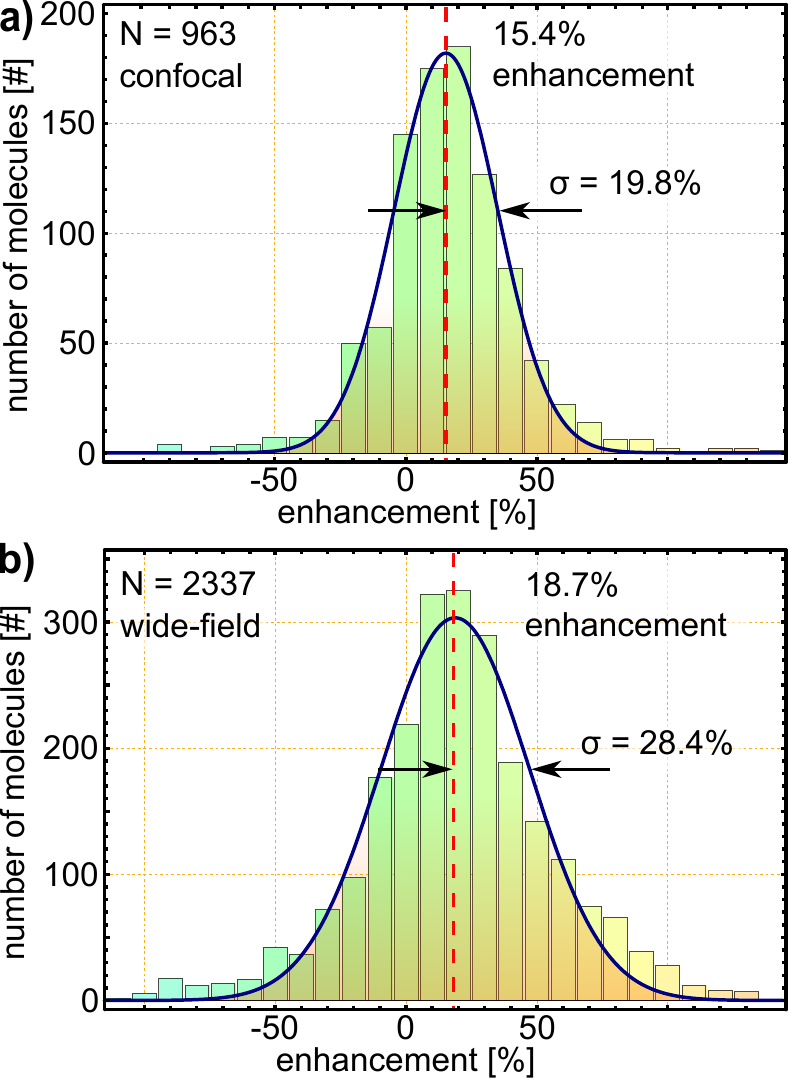}
  \caption{Direct comparison of the enhancement of multiple molecules in the different filter configurations. Molecules tend to blink and bleach, such that statistical data processing is necessary. a) 963 molecules, detected in a confocal configuration. The signal for the atomic notch filter is improved by about 15\% for the atomic notch filter. N=number of molecules. b) 2337 molecules, detected in a wide-field microscope. The enhancement is shown to be about 18\%. The statistical distribution of the enhancement is wider than in the confocal case due to the generally lower SNR than in the confocal configuration.}
  \label{fig:comparison}
\end{figure}

In an statistical analysis of 963 molecules, when each molecule is compared in both filter configurations, we see an enhancement in the overall detected counts per emitter of 15.4\%. Fig.~\ref{fig:comparison}a shows a histogram for all recorded molecules. The statistical error defined as $\sigma / \sqrt{N}$, is $1.5$\%. The atomic filter therefore increases the number of collected photons, but unfortunately the background suppression is one order of magnitude smaller for the vapor cell to the commercial filter.

\subsection*{Wide-field Imaging}
In the wide-field experiment no pinhole is introduced in contrast to the confocal setup. According to that, clipping of a wandering beam is not critical. Instead, convection of hot air originating from the vapor cell leads to a shifted or blurred image. In fact, the image is found slightly blurred, and, in addition, the lateral extension of the molecules is not diffraction limited anymore. The mean width of the fluorescence molecules for the atomic cell is $427$~nm and for the commercial filter $365$~nm. In both cases the statistical error is more than two orders of magnitude smaller. The resolution is clearly reduced against the confocal configuration.

As in the confocal case, the images look comparable between the two filter configurations (Fig.~\ref{fig:widefield}a and b). The background contribution in the atomic filter case is increased, but not as significant as in the confocal configuration. In an analysis of all acquired images, the mean background contribution is increased by 30\% from 70 to 100~cps. This increase is also visible in the line-cut shown in Fig.~\ref{fig:widefield}c.

Wide-field imaging shows further an only slightly higher background fluctuation, which is likely caused by a limited depth selection of the objective focal plane. As for the integrated signal, an enhancement of $18.7$\%  with a standard deviation of $\sigma=28.4$\% is observed. Taking $2337$ measured molecules into account, this leads to a statistical uncertainty of 0.6\%. The noise level of the commercial filter and sodium filter is determined to 30 vs.\ 40 photon per seconds, based on a laser excitation of 50~$\mu$W into the microscope. The increased noise of the atomic filter automatically lowers the SNR. However, the higher signal not only compensates this drawback, but increases the SNR. We find an SNR of 60 for the atomic filtering vs.\ 31 with the commercial filter. The intrinsic noise level of the camera is one order of magnitude less and does not play a role here. In summary, we achieve an increased signal, as well as an increased signal to noise ratio for wide-field imaging with the atomic filter configuration.

\subsection*{Overall Enhancement}

Fig.~\ref{fig:comparison} shows a histogram of the determined integrated count rates by the commercial and the atomic filtering schemes. Molecules tend to blink and bleach, and it is possible that a molecule was fully bright in one image and much dimmer in the corresponding image, which have been acquired with the other filter. The higher fluctuation on the camera (wandering image) leads to an increased spread of the resulting count rate. Therefore, a much wider distribution than in the confocal case is observed. Another important factor is the signal to noise ratio (SNR) in the analyzed image. When e.g.\ the background noise in the image is higher, the resulting fit outcome for a single molecule signal shows higher fluctuations. Thereby, Fig.~\ref{fig:comparison} does not only represent the overall enhancement of the signal, but also represents the SNR.

In summary, the confocal images show a lower SNR and SBR for an image acquisition with an atomic vapor cell. But the lateral spread of the molecules is not changing, and we find an enhancement in the number detected photons of 15.4$\pm$1.5~\%.

The wide-field images show a comparable SBR. For wide-field applications the vapor-cell filter exhibits a better performance than the commercial filter of about 18~$\pm$0.6\% for the signal and a by factor two increased SNR.

The atomic filtering results in an enhancement of detectable photons on the order of 15\% in both configurations. This is fully consistent with the acquisition of the single molecule spectra as shown in Fig.~\ref{fig:spectra}a, which also showed an enhancement of approximately 15\%.

\section*{Conclusion and Outlook}

An experimental alternative to dichroic filters or commercial notch filters was presented. It allows for studies on low fluorescing samples down to the single molecule level. The enhancement of about 15\% in the detected photon number can be a crucial enhancement for microscopy and sensing. The increased background fluctuations in the confocal case need to be addressed, eventually with a widened beam to lower the intensity inside the vapor cell. Also the use of a Fabry-P\'{e}rot cavity to clean up the excitation beam from the fiber auto-fluorescent should be used. In Raman spectroscopy atom-based narrow-band filters were explored in the near-infrared part of the spectrum~\cite{lin_ol_2014}, but such experiments were not performed with atomic sodium and yellow light. We underline that the introduced filter matches well to many dyes which are used in (micro-)biological imaging. So far, only other vapor cells, such as Cs, Rb, have been used. Such filters can allow for the sensitive detection also of low-fluorescing defect centers, such as defects in silicon-carbide~\cite{widmann_nmat_2015}, which could be eventually combined e.g.\ with atomic rubidium or cesium.

The experimental challenges are not necessarily easy to solve in a number of micro-biology labs. The introduced filtering option allows for enhancing the overall collection efficiency, if everything else has been optimized. A convenient complement to this notch filter is the use of a Faraday anomalous dispersion optical filter (FADOF), which represents a GHz-wide band-pass configuration~\cite{dick_ol_1991,kiefer_srep_2014}. In the future, especially, when convenient diode lasers locked to the sodium wavelength are available, the technique of filtering a single molecule experiment by an atomic vapor might be giving a small, but crucial enhancement of detection efficiency, which allows for better sensing and localization accuracies in material-science and applications in micro-biology.

\begin{backmatter}

\section*{Competing interests}
  The authors declare that they have no competing interests.

\section*{Author's contributions}
IG envisioned and prepared the experiments. TR helped with the configuration and the design of the microscope. DU and TR performed the measurements. IG, DU, TR, and MW processed the data. JW, SL and IG supervised the team. IG, MW and TR wrote the manuscript.

\section*{Acknowledgments}
  We thank Andrea Zappe for DNA preparation. We also thank Helmut Kammerlander from the glass shop for producing several sodium cells. Initial experiments were conducted by Guilherme Stein. We acknowledge funding from the Max Planck Society (JW, Max Planck fellowship).


\newcommand{\BMCxmlcomment}[1]{}

\BMCxmlcomment{

<refgrp>

<bibl id="B1">
  <title><p>Optical microscopic observation of single small
  molecules</p></title>
  <aug>
    <au><snm>Hirschfeld</snm><fnm>T.</fnm></au>
  </aug>
  <source>Appl. Opt.</source>
  <publisher>OSA</publisher>
  <pubdate>1976</pubdate>
  <volume>15</volume>
  <issue>12</issue>
  <fpage>2965</fpage>
  <lpage>-2966</lpage>
  <url>http://ao.osa.org/abstract.cfm?URI=ao-15-12-2965</url>
</bibl>

<bibl id="B2">
  <title><p>{Optical detection and spectroscopy of single molecules in a
  solid}</p></title>
  <aug>
    <au><snm>Moerner</snm><fnm>W</fnm></au>
    <au><snm>Kador</snm><fnm>L</fnm></au>
  </aug>
  <source>Physical Review Letters</source>
  <pubdate>1989</pubdate>
  <volume>62</volume>
  <issue>21</issue>
  <fpage>2535</fpage>
  <lpage>-2538</lpage>
</bibl>

<bibl id="B3">
  <title><p>Single pentacene molecules detected by fluorescence excitation in a
  \textit{p}-terphenyl crystal</p></title>
  <aug>
    <au><snm>Orrit</snm><fnm>M.</fnm></au>
    <au><snm>Bernard</snm><fnm>J.</fnm></au>
  </aug>
  <source>Phys. Rev. Lett.</source>
  <publisher>American Physical Society</publisher>
  <pubdate>1990</pubdate>
  <volume>65</volume>
  <fpage>2716</fpage>
  <lpage>-2719</lpage>
  <url>http://link.aps.org/doi/10.1103/PhysRevLett.65.2716</url>
</bibl>

<bibl id="B4">
  <title><p>{Confocal microscopy with an increased detection aperture: type-B
  4Pi confocal microscopy}</p></title>
  <aug>
    <au><snm>Hell</snm><fnm>S.\ {}W.</fnm></au>
    <au><snm>Stelzer</snm><fnm>E.\ {}H.\ {}K.</fnm></au>
    <au><snm>Lindek</snm><fnm>S</fnm></au>
    <au><snm>Cremer</snm><fnm>C</fnm></au>
  </aug>
  <source>Optics Letters</source>
  <pubdate>1994</pubdate>
  <volume>19</volume>
  <fpage>222</fpage>
  <lpage>-224</lpage>
</bibl>

<bibl id="B5">
  <title><p>Imaging Intracellular Fluorescent Proteins at Nanometer
  Resolution</p></title>
  <aug>
    <au><snm>Betzig</snm><fnm>E</fnm></au>
    <au><snm>Patterson</snm><fnm>GH</fnm></au>
    <au><snm>Sougrat</snm><fnm>R</fnm></au>
    <au><snm>Lindwasser</snm><fnm>OW</fnm></au>
    <au><snm>Olenych</snm><fnm>S</fnm></au>
    <au><snm>Bonifacino</snm><fnm>JS</fnm></au>
    <au><snm>Davidson</snm><fnm>MW</fnm></au>
    <au><snm>Lippincott Schwartz</snm><fnm>J</fnm></au>
    <au><snm>Hess</snm><fnm>HF</fnm></au>
  </aug>
  <source>Science</source>
  <pubdate>2006</pubdate>
  <volume>313</volume>
  <issue>5793</issue>
  <fpage>1642</fpage>
  <lpage>1645</lpage>
</bibl>

<bibl id="B6">
  <title><p>Sub-diffraction-limit imaging by stochastic optical reconstruction
  microscopy (STORM)</p></title>
  <aug>
    <au><snm>Rust</snm><fnm>MJ</fnm></au>
    <au><snm>Bates</snm><fnm>M</fnm></au>
    <au><snm>Zhuang</snm><fnm>X</fnm></au>
  </aug>
  <source>Nat Meth</source>
  <pubdate>2006</pubdate>
  <volume>3</volume>
  <issue>10</issue>
  <fpage>793</fpage>
  <lpage>-796</lpage>
  <url>http://dx.doi.org/10.1038/nmeth929</url>
</bibl>

<bibl id="B7">
  <title><p>Transcription by single molecules of {RNA} polymerase observed by
  light microscopy.</p></title>
  <aug>
    <au><snm>Schafer</snm><fnm>DA</fnm></au>
    <au><snm>Gelles</snm><fnm>J</fnm></au>
    <au><snm>Sheetz</snm><fnm>MP</fnm></au>
    <au><snm>Landick</snm><fnm>R</fnm></au>
  </aug>
  <source>Nature</source>
  <pubdate>1991</pubdate>
  <volume>352</volume>
  <issue>6334</issue>
  <fpage>444</fpage>
  <lpage>-448</lpage>
</bibl>

<bibl id="B8">
  <title><p>Probing individual molecules with confocal fluorescence
  microscopy</p></title>
  <aug>
    <au><snm>Nie</snm><fnm>S</fnm></au>
    <au><snm>Chiu</snm><fnm>DT</fnm></au>
    <au><snm>Zare</snm><fnm>RN</fnm></au>
  </aug>
  <source>Science</source>
  <pubdate>1994</pubdate>
  <volume>266</volume>
  <issue>5187</issue>
  <fpage>1018</fpage>
  <lpage>1021</lpage>
  <url>http://www.sciencemag.org/content/266/5187/1018.abstract</url>
</bibl>

<bibl id="B9">
  <title><p>Optical detection of single molecules</p></title>
  <aug>
    <au><snm>Nie</snm><fnm>S</fnm></au>
    <au><snm>Zare</snm><fnm>RN</fnm></au>
  </aug>
  <source>Annual review of biophysics and biomolecular structure</source>
  <publisher>Annual Reviews 4139 El Camino Way, PO Box 10139, Palo Alto, CA
  94303-0139, USA</publisher>
  <pubdate>1997</pubdate>
  <volume>26</volume>
  <issue>1</issue>
  <fpage>567</fpage>
  <lpage>-596</lpage>
</bibl>

<bibl id="B10">
  <title><p>Fluorescence spectroscopy of single biomolecules</p></title>
  <aug>
    <au><snm>Weiss</snm><fnm>S</fnm></au>
  </aug>
  <source>Science</source>
  <publisher>American Association for the Advancement of Science</publisher>
  <pubdate>1999</pubdate>
  <volume>283</volume>
  <issue>5408</issue>
  <fpage>1676</fpage>
  <lpage>-1683</lpage>
</bibl>

<bibl id="B11">
  <title><p>{Single-molecule high-resolution imaging with
  photobleaching}</p></title>
  <aug>
    <au><snm>Gordon</snm><fnm>MP</fnm></au>
    <au><snm>Ha</snm><fnm>T</fnm></au>
    <au><snm>Selvin</snm><fnm>PR</fnm></au>
  </aug>
  <source>Proceedings of the National Academy of Sciences of the United States
  of America</source>
  <pubdate>2004</pubdate>
  <volume>101</volume>
  <issue>17</issue>
  <fpage>6462</fpage>
  <lpage>-6465</lpage>
  <url>http://www.pnas.org/content/101/17/6462.abstract</url>
</bibl>

<bibl id="B12">
  <title><p>A planar dielectric antenna for directional single-photon emission
  and near-unity collection efficiency</p></title>
  <aug>
    <au><snm>Lee</snm><fnm>K. G.</fnm></au>
    <au><snm>Chen</snm><fnm>W. X.</fnm></au>
    <au><snm>Eghlidi</snm><fnm>H.</fnm></au>
    <au><snm>Kukura</snm><fnm>P.</fnm></au>
    <au><snm>Lettow</snm><fnm>R.</fnm></au>
    <au><snm>Renn</snm><fnm>A.</fnm></au>
    <au><snm>Sandoghdar</snm><fnm>V.</fnm></au>
    <au><snm>G\"{o}tzinger</snm><fnm>S.</fnm></au>
  </aug>
  <source>Nature Photonics</source>
  <publisher>Nature Publishing Group</publisher>
  <pubdate>2011</pubdate>
  <volume>5</volume>
  <issue>3</issue>
  <fpage>166</fpage>
  <lpage>-169</lpage>
  <url>http://dx.doi.org/10.1038/nphoton.2010.312</url>
</bibl>

<bibl id="B13">
  <title><p>{Spectroscopy of single N-V centers in diamond}</p></title>
  <aug>
    <au><snm>Jelezko</snm><fnm>F</fnm></au>
    <au><snm>Tietz</snm><fnm>C</fnm></au>
    <au><snm>Gruber</snm><fnm>A.</fnm></au>
    <au><snm>Popa</snm><fnm>I</fnm></au>
    <au><snm>Nizovtsev</snm><fnm>A.</fnm></au>
    <au><snm>Kilin</snm><fnm>S</fnm></au>
    <au><snm>Wrachtrup</snm><fnm>J</fnm></au>
  </aug>
  <source>Single Molecules</source>
  <pubdate>2001</pubdate>
  <volume>2</volume>
  <fpage>255</fpage>
  <lpage>-260</lpage>
</bibl>

<bibl id="B14">
  <title><p>{Molecular-sized fluorescent nanodiamonds}</p></title>
  <aug>
    <au><snm>Vlasov</snm><fnm>II</fnm></au>
    <au><snm>Shiryaev</snm><fnm>AA</fnm></au>
    <au><snm>Rendler</snm><fnm>T</fnm></au>
    <au><snm>Steinert</snm><fnm>S</fnm></au>
    <au><snm>Lee</snm><fnm>SY</fnm></au>
    <au><snm>Antonov</snm><fnm>D</fnm></au>
    <au><snm>Voros</snm><fnm>M</fnm></au>
    <au><snm>Jelezko</snm><fnm>F</fnm></au>
    <au><snm>Fisenko</snm><fnm>AV</fnm></au>
    <au><snm>Semjonova</snm><fnm>LF</fnm></au>
    <au><snm>Biskupek</snm><fnm>J</fnm></au>
    <au><snm>Kaiser</snm><fnm>U</fnm></au>
    <au><snm>Lebedev</snm><fnm>OI</fnm></au>
    <au><snm>Sildos</snm><fnm>I</fnm></au>
    <au><snm>Hemmer</snm><fnm>PR</fnm></au>
    <au><snm>Konov</snm><fnm>VI</fnm></au>
    <au><snm>Gali</snm><fnm>A</fnm></au>
    <au><snm>Wrachtrup</snm><fnm>J</fnm></au>
  </aug>
  <source>Nature Nanotechnology</source>
  <publisher>Nature Publishing Group</publisher>
  <pubdate>2014</pubdate>
  <volume>9</volume>
  <issue>1</issue>
  <fpage>54</fpage>
  <lpage>-58</lpage>
  <url>http://dx.doi.org/10.1038/nnano.2013.255 10.1038/nnano.2013.255
  http://www.nature.com/nnano/journal/v9/n1/abs/nnano.2013.255.html\#supplementary-information</url>
</bibl>

<bibl id="B15">
  <title><p>Semiconductor quantum dots and metal nanoparticles: syntheses,
  optical properties, and biological applications</p></title>
  <aug>
    <au><snm>Biju</snm><fnm>V</fnm></au>
    <au><snm>Itoh</snm><fnm>T</fnm></au>
    <au><snm>Anas</snm><fnm>A</fnm></au>
    <au><snm>Sujith</snm><fnm>A</fnm></au>
    <au><snm>Ishikawa</snm><fnm>M</fnm></au>
  </aug>
  <source>Analytical and Bioanalytical Chemistry</source>
  <publisher>Springer-Verlag</publisher>
  <pubdate>2008</pubdate>
  <volume>391</volume>
  <issue>7</issue>
  <fpage>2469</fpage>
  <lpage>2495</lpage>
  <url>http://dx.doi.org/10.1007/s00216-008-2185-7</url>
</bibl>

<bibl id="B16">
  <title><p>{The new fluorescent probes on the block}</p></title>
  <aug>
    <au><snm>Evanko</snm><fnm>D</fnm></au>
  </aug>
  <source>Nature Methods</source>
  <publisher>Nature Publishing Group</publisher>
  <pubdate>2008</pubdate>
  <volume>5</volume>
  <issue>3</issue>
  <fpage>218</fpage>
  <lpage>-219</lpage>
  <url>http://dx.doi.org/10.1038/nmeth0308-218a</url>
</bibl>

<bibl id="B17">
  <title><p>Microscopic diamond solid-immersion-lenses fabricated around single
  defect centers by focused ion beam milling</p></title>
  <aug>
    <au><snm>Jamali</snm><fnm>M</fnm></au>
    <au><snm>Gerhardt</snm><fnm>I</fnm></au>
    <au><snm>Rezai</snm><fnm>M</fnm></au>
    <au><snm>Frenner</snm><fnm>K</fnm></au>
    <au><snm>Fedder</snm><fnm>H</fnm></au>
    <au><snm>Wrachtrup</snm><fnm>J</fnm></au>
  </aug>
  <source>Review of Scientific Instruments</source>
  <pubdate>2014</pubdate>
  <volume>85</volume>
  <issue>12</issue>
  <url>http://scitation.aip.org/content/aip/journal/rsi/85/12/10.1063/1.4902818</url>
</bibl>

<bibl id="B18">
  <title><p>{Spectra of the D-Lines of Alkali Vapors}</p></title>
  <aug>
    <au><snm>Domenico</snm><fnm>GD</fnm></au>
    <au><snm>Weis</snm><fnm>A</fnm></au>
  </aug>
  <source>Wolfram Demonstrations Project</source>
  <pubdate>2011</pubdate>
  <url>http://demonstrations.wolfram.com/SpectraOfTheDLinesOfAlkaliVapors/</url>
</bibl>

<bibl id="B19">
  <title><p>Super efficient absorption filter for quantum memory using atomic
  ensembles in a vapor</p></title>
  <aug>
    <au><snm>Heifetz</snm><fnm>A</fnm></au>
    <au><snm>Agarwal</snm><fnm>A</fnm></au>
    <au><snm>Cardoso</snm><fnm>GC</fnm></au>
    <au><snm>Gopal</snm><fnm>V</fnm></au>
    <au><snm>Kumar</snm><fnm>P</fnm></au>
    <au><snm>Shahriar</snm><fnm>M.S.</fnm></au>
  </aug>
  <source>Optics Communications</source>
  <pubdate>2004</pubdate>
  <volume>232</volume>
  <issue>1?6</issue>
  <fpage>289</fpage>
  <lpage>293</lpage>
  <url>http://www.sciencedirect.com/science/article/pii/S0030401804000070</url>
</bibl>

<bibl id="B20">
  <title><p>{Absolute absorption and dispersion of a rubidium vapour in the
  hyperfine Paschen–Back regime}</p></title>
  <aug>
    <au><snm>Weller</snm><fnm>L</fnm></au>
    <au><snm>Kleinbach</snm><fnm>KS</fnm></au>
    <au><snm>Zentile</snm><fnm>MA</fnm></au>
    <au><snm>Knappe</snm><fnm>S</fnm></au>
    <au><snm>Adams</snm><fnm>CS</fnm></au>
    <au><snm>Hughes</snm><fnm>IG</fnm></au>
  </aug>
  <source>Journal of Physics B: Atomic, Molecular and Optical Physics</source>
  <pubdate>2012</pubdate>
  <volume>45</volume>
  <issue>21</issue>
  <fpage>215005</fpage>
  <url>http://stacks.iop.org/0953-4075/45/i=21/a=215005</url>
</bibl>

<bibl id="B21">
  <title><p>The rubidium atomic clock and basic research</p></title>
  <aug>
    <au><snm>Camparo</snm><fnm>J</fnm></au>
  </aug>
  <source>Physics today</source>
  <publisher>American Institute of Physics</publisher>
  <pubdate>2007</pubdate>
  <volume>60</volume>
  <issue>11</issue>
  <fpage>33</fpage>
  <lpage>-39</lpage>
</bibl>

<bibl id="B22">
  <title><p>{Ultraviolet Raman spectroscopy using an atomic vapor filter and
  incoherent excitation}</p></title>
  <aug>
    <au><snm>Pelletier</snm><fnm>M J</fnm></au>
  </aug>
  <source>Applied spectroscopy</source>
  <publisher>Society for Applied Spectroscopy</publisher>
  <pubdate>1992</pubdate>
  <volume>46</volume>
  <issue>3</issue>
  <fpage>395</fpage>
  <lpage>-400</lpage>
</bibl>

<bibl id="B23">
  <title><p>{Daytime mesopause temperature measurements with a sodium-vapor
  dispersive Faraday filter in a lidar receiver.}</p></title>
  <aug>
    <au><snm>Chen</snm><fnm>H</fnm></au>
    <au><snm>White</snm><fnm>Ma</fnm></au>
    <au><snm>Krueger</snm><fnm>Da</fnm></au>
    <au><snm>She</snm><fnm>C Y</fnm></au>
  </aug>
  <source>Optics Letters</source>
  <pubdate>1996</pubdate>
  <volume>21</volume>
  <issue>15</issue>
  <fpage>1093</fpage>
  <lpage>-1095</lpage>
</bibl>

<bibl id="B24">
  <title><p>{Experimental study of a model digital space optical communication
  system with new quantum devices.}</p></title>
  <aug>
    <au><snm>Junxiong</snm><fnm>T</fnm></au>
    <au><snm>Qingji</snm><fnm>W</fnm></au>
    <au><snm>Yimin</snm><fnm>L</fnm></au>
    <au><snm>Liang</snm><fnm>Z</fnm></au>
    <au><snm>Jianhua</snm><fnm>G</fnm></au>
    <au><snm>Minghao</snm><fnm>D</fnm></au>
    <au><snm>Jiankun</snm><fnm>K</fnm></au>
    <au><snm>Lemin</snm><fnm>Z</fnm></au>
  </aug>
  <source>Applied Optics</source>
  <pubdate>1995</pubdate>
  <volume>34</volume>
  <issue>15</issue>
  <fpage>2619</fpage>
  <lpage>-2622</lpage>
</bibl>

<bibl id="B25">
  <title><p>{Ultralow frequency Stokes and anti-Stokes Raman spectroscopy of
  single living cells and microparticles using a hot rubidium vapor
  filter}</p></title>
  <aug>
    <au><snm>Lin</snm><fnm>J</fnm></au>
    <au><snm>Li</snm><fnm>Yq</fnm></au>
  </aug>
  <source>Optics Letters</source>
  <publisher>OSA</publisher>
  <pubdate>2014</pubdate>
  <volume>39</volume>
  <issue>1</issue>
  <fpage>108</fpage>
  <lpage>-110</lpage>
  <url>http://ol.osa.org/abstract.cfm?URI=ol-39-1-108</url>
</bibl>

<bibl id="B26">
  <title><p>{Microscopy apparatus}</p></title>
  <aug>
    <au><snm>Minsky</snm><fnm>M</fnm></au>
  </aug>
  <pubdate>1961</pubdate>
  <url>http://worldwide.espacenet.com/publicationDetails/biblio?CC=US\&NR=3013467\&KC=\&FT=E\&locale=en\_EP</url>
</bibl>

<bibl id="B27">
  <title><p>{A Microscope for Observation of Fluorescence in Living
  Tissues}</p></title>
  <aug>
    <au><snm>Singer</snm><fnm>E</fnm></au>
  </aug>
  <source>Science</source>
  <pubdate>1932</pubdate>
  <volume>75</volume>
  <issue>1941</issue>
  <fpage>289</fpage>
  <lpage>-291</lpage>
  <url>http://www.sciencemag.org/content/75/1941/289.2.short</url>
</bibl>

<bibl id="B28">
  <title><p>{Molecular photons interfaced with alkali atoms}</p></title>
  <aug>
    <au><snm>Siyushev</snm><fnm>P</fnm></au>
    <au><snm>Stein</snm><fnm>G</fnm></au>
    <au><snm>Wrachtrup</snm><fnm>J</fnm></au>
    <au><snm>Gerhardt</snm><fnm>I</fnm></au>
  </aug>
  <source>Nature</source>
  <pubdate>2014</pubdate>
  <volume>509</volume>
  <issue>7498</issue>
  <fpage>66</fpage>
  <lpage>-70</lpage>
  <url>http://dx.doi.org/10.1038/nature13191</url>
</bibl>

<bibl id="B29">
  <title><p>{Doppler-free spectroscopy using magnetically induced dichroism of
  atomic vapor: a new scheme for laser frequency locking}</p></title>
  <aug>
    <au><snm>Petelski</snm><fnm>T</fnm></au>
    <au><snm>Fattori</snm><fnm>M</fnm></au>
    <au><snm>Lamporesi</snm><fnm>G</fnm></au>
    <au><snm>Stuhler</snm><fnm>J</fnm></au>
    <au><snm>Tino</snm><fnm>G M</fnm></au>
  </aug>
  <source>The European Physical Journal D - Atomic, Molecular, Optical and
  Plasma Physics</source>
  <publisher>EDP Sciences, Springer-Verlag, Societ\`{a} Italiana di
  Fisica</publisher>
  <pubdate>2003</pubdate>
  <volume>22</volume>
  <issue>2</issue>
  <fpage>279</fpage>
  <lpage>-283</lpage>
  <url>http://dx.doi.org/10.1140/epjd/e2002-00238-4</url>
</bibl>

<bibl id="B30">
  <title><p>A sodium-resistant glass cell by coating with CaF 2</p></title>
  <aug>
    <au><snm>Laux</snm><fnm>L</fnm></au>
    <au><snm>Schulz</snm><fnm>G</fnm></au>
  </aug>
  <source>Journal of Physics E: Scientific Instruments</source>
  <pubdate>1980</pubdate>
  <volume>13</volume>
  <issue>8</issue>
  <fpage>823</fpage>
  <url>http://stacks.iop.org/0022-3735/13/i=8/a=006</url>
</bibl>

<bibl id="B31">
  <title><p>Construction of long-life magneto-optical filters for
  helioseismology observations</p></title>
  <aug>
    <au><snm>Sakurai</snm><fnm>T.</fnm></au>
    <au><snm>Tanaka</snm><fnm>K.</fnm></au>
    <au><snm>Miyazaki</snm><fnm>H.</fnm></au>
    <au><snm>Ichimoto</snm><fnm>K.</fnm></au>
    <au><snm>Sakata</snm><fnm>A.</fnm></au>
    <au><snm>Wada</snm><fnm>S.</fnm></au>
  </aug>
  <source>Progress of Seismology of the Sun and Stars</source>
  <publisher>Springer Berlin Heidelberg</publisher>
  <editor>Osaki, Y. and Shibahashi, H.</editor>
  <series><title><p>Lecture Notes in Physics</p></title></series>
  <pubdate>1990</pubdate>
  <volume>367</volume>
  <fpage>277</fpage>
  <lpage>280</lpage>
</bibl>

<bibl id="B32">
  <title><p>Fluorescence quantum yields of some rhodamine dyes</p></title>
  <aug>
    <au><snm>Kubin</snm><fnm>R.F.</fnm></au>
    <au><snm>Fletcher</snm><fnm>A.N.</fnm></au>
  </aug>
  <source>Journal of Luminescence</source>
  <pubdate>1983</pubdate>
  <volume>27</volume>
  <issue>4</issue>
  <fpage>455</fpage>
  <lpage>462</lpage>
  <url>http://www.sciencedirect.com/science/article/pii/002223138290045X</url>
</bibl>

<bibl id="B33">
  <title><p>Influence of the molecular structure and the nature of the solvent
  on the absorption and fluorescence characteristics of rhodamines</p></title>
  <aug>
    <au><snm>Arbeloa</snm><fnm>FL</fnm></au>
    <au><snm>Aguirresacona</snm><fnm>I</fnm></au>
    <au><snm>Arbeloa</snm><fnm>I</fnm></au>
  </aug>
  <source>Chemical Physics</source>
  <pubdate>1989</pubdate>
  <volume>130</volume>
  <issue>1-3</issue>
  <fpage>371</fpage>
  <lpage>378</lpage>
  <url>http://www.sciencedirect.com/science/article/pii/0301010489870661</url>
</bibl>

<bibl id="B34">
  <title><p>{Coherent control of single spins in silicon carbide at room
  temperature}</p></title>
  <aug>
    <au><snm>Widmann</snm><fnm>M</fnm></au>
    <au><snm>Lee</snm><fnm>SY</fnm></au>
    <au><snm>Rendler</snm><fnm>T</fnm></au>
    <au><snm>Son</snm><fnm>NT</fnm></au>
    <au><snm>Fedder</snm><fnm>H</fnm></au>
    <au><snm>Paik</snm><fnm>S</fnm></au>
    <au><snm>Yang</snm><fnm>LP</fnm></au>
    <au><snm>Zhao</snm><fnm>N</fnm></au>
    <au><snm>Yang</snm><fnm>S</fnm></au>
    <au><snm>Booker</snm><fnm>I</fnm></au>
    <au><snm>Denisenko</snm><fnm>A</fnm></au>
    <au><snm>Jamali</snm><fnm>M</fnm></au>
    <au><snm>Momenzadeh</snm><fnm>SA</fnm></au>
    <au><snm>Gerhardt</snm><fnm>I</fnm></au>
    <au><snm>Ohshima</snm><fnm>T</fnm></au>
    <au><snm>Gali</snm><fnm>A</fnm></au>
    <au><snm>Janz\'{e}n</snm><fnm>E</fnm></au>
    <au><snm>Wrachtrup</snm><fnm>J</fnm></au>
  </aug>
  <source>Nature Materials</source>
  <publisher>Nature Publishing Group</publisher>
  <pubdate>2015</pubdate>
  <volume>14</volume>
  <issue>2</issue>
  <fpage>164</fpage>
  <lpage>-168</lpage>
  <url>http://dx.doi.org/10.1038/nmat4145 10.1038/nmat4145
  http://www.nature.com/nmat/journal/v14/n2/abs/nmat4145.html\#supplementary-information</url>
</bibl>

<bibl id="B35">
  <title><p>Ultrahigh-noise rejection optical filter</p></title>
  <aug>
    <au><snm>Dick</snm><fnm>D. J.</fnm></au>
    <au><snm>Shay</snm><fnm>T. M.</fnm></au>
  </aug>
  <source>Opt. Lett.</source>
  <publisher>OSA</publisher>
  <pubdate>1991</pubdate>
  <volume>16</volume>
  <issue>11</issue>
  <fpage>867</fpage>
  <lpage>-869</lpage>
  <url>http://ol.osa.org/abstract.cfm?URI=ol-16-11-867</url>
</bibl>

<bibl id="B36">
  <title><p>{Na-Faraday rotation filtering: The optimal point}</p></title>
  <aug>
    <au><snm>Kiefer</snm><fnm>W</fnm></au>
    <au><snm>L\"{o}w</snm><fnm>R</fnm></au>
    <au><snm>Wrachtrup</snm><fnm>J</fnm></au>
    <au><snm>Gerhardt</snm><fnm>I</fnm></au>
  </aug>
  <source>Scientific Reports</source>
  <publisher>Macmillan Publishers Limited. All rights reserved</publisher>
  <pubdate>2014</pubdate>
  <volume>4</volume>
  <url>http://dx.doi.org/10.1038/srep06552 10.1038/srep06552
  http://www.nature.com/srep/2014/141009/srep06552/abs/srep06552.html\#supplementary-information</url>
</bibl>

</refgrp>
} 

\end{backmatter}
\end{document}